\documentclass[onecolumn,epjc3]{svjour3} 
 
\usepackage[T1]{fontenc}
\usepackage{epsfig,amsmath,amssymb}
\usepackage{graphicx}
\usepackage{xspace}
\usepackage{listings}
\usepackage{ulem}
\usepackage{slashed}
\usepackage{cite}
\usepackage{xcolor}
\usepackage{nicefrac}
\usepackage{feynmp}
\usepackage[colorlinks,citecolor=blue,urlcolor=blue,linkcolor=blue]{hyperref}
\journalname{Eur. Phys. J. C}

\definecolor{maroon}{cmyk}{0, 0.87, 0.68, 0.32}
\definecolor{halfgray}{gray}{0.55}
\definecolor{slha_frame}{RGB}{207, 207, 207}
\definecolor{slha_bg}{RGB}{247, 247, 247}
\definecolor{slha_red}{RGB}{186, 33, 33}
\definecolor{slha_green}{RGB}{0, 128, 0}
\definecolor{slha_cyan}{RGB}{64, 128, 128}
\definecolor{slha_purple}{RGB}{170, 34, 255}

\definecolor{mathematica_frame}{RGB}{207, 207, 207}
\definecolor{mathematica_bg}{RGB}{247, 247, 247}
\definecolor{mathematica_red}{RGB}{186, 33, 33}
\definecolor{mathematica_green}{RGB}{0, 128, 0}
\definecolor{mathematica_cyan}{RGB}{64, 128, 128}
\definecolor{mathematica_purple}{RGB}{170, 34, 255}

\lstnewenvironment{MIN}[1][]{%
  \renewcommand{\thelstnumber}{In[\arabic{lstnumber}]}
  \lstset{language=MathIn,numbers=left,basicstyle=\ttfamily,#1}%
}{%
}

\lstnewenvironment{MOUT}[1][]{%
  \renewcommand{\thelstnumber}{Out[\arabic{lstnumber}]}
  \lstset{language=MathOut,numbers=left,basicstyle=\ttfamily,#1}%
}{%
}

\usepackage{listings}
\lstdefinelanguage{SLHA}{
    morekeywords={block,Block,BLOCK,decay,Decay,DECAY},%
    sensitive=true,%
    morecomment=[l]\#,%
    morestring=[b]',%
    morestring=[b]",%
    morestring=[s]{'''}{'''},
    morestring=[s]{"""}{"""},
    morestring=[s]{r'}{'},
    morestring=[s]{r"}{"},%
    morestring=[s]{r'''}{'''},%
    morestring=[s]{r"""}{"""},%
    morestring=[s]{u'}{'},
    morestring=[s]{u"}{"},%
    morestring=[s]{u'''}{'''},%
    morestring=[s]{u"""}{"""},%
    identifierstyle=\color{black}\ttfamily,
    commentstyle=\color{slha_cyan}\ttfamily,
    stringstyle=\color{slha_red}\ttfamily,
    keepspaces=true,
    showspaces=false,
    showstringspaces=false,
    rulecolor=\color{slha_frame},
    frame=single,
    frameround={t}{t}{t}{t},
    framexleftmargin=6mm,
    numbers=left,
    numberstyle=\tiny\color{halfgray},
    backgroundcolor=\color{slha_bg},
    basicstyle=\footnotesize,
    keywordstyle=\color{slha_green}\ttfamily,
    aboveskip=1.2em,
    belowskip=1.2em,
}

\lstdefinelanguage{MathIn}{
    morekeywords={Simplify,Eigenvalues},%
    emph={Start,InitUnitarity,GetScatteringDiagrams,BuildScatteringMatrix,MakeSPheno},%
    emphstyle={\color{mathematica_purple}},
    sensitive=true,%
    morecomment=[l]\%,%
    morestring=[b]',%
    morestring=[b]",%
    morestring=[s]{'''}{'''},
    morestring=[s]{"""}{"""},
    morestring=[s]{r'}{'},
    morestring=[s]{r"}{"},%
    morestring=[s]{r'''}{'''},%
    morestring=[s]{r"""}{"""},%
    morestring=[s]{u'}{'},
    morestring=[s]{u"}{"},%
    morestring=[s]{u'''}{'''},%
    morestring=[s]{u"""}{"""},%
    identifierstyle=\color{black}\ttfamily,
    commentstyle=\color{mathematica_cyan}\ttfamily,
    stringstyle=\color{mathematica_red}\ttfamily,
    keepspaces=true,
    showspaces=false,
    showstringspaces=false,
    rulecolor=\color{mathematica_frame},
    frame=single,
    frameround={t}{t}{t}{t},
    framexleftmargin=10mm,
    numbers=left,
    numberstyle=\tiny\color{halfgray},
    backgroundcolor=\color{mathematica_bg},
    basicstyle=\footnotesize,
    keywordstyle=\color{mathematica_green}\ttfamily,
    aboveskip=1.2em,
    belowskip=1.2em,
}

\lstdefinelanguage{MathOut}{
    morekeywords={Simplify,Eigenvalues},%
    sensitive=true,%
    morecomment=[l]\%,%
    morestring=[b]',%
    morestring=[b]",%
    morestring=[s]{'''}{'''},
    morestring=[s]{"""}{"""},
    morestring=[s]{r'}{'},
    morestring=[s]{r"}{"},%
    morestring=[s]{r'''}{'''},%
    morestring=[s]{r"""}{"""},%
    morestring=[s]{u'}{'},
    morestring=[s]{u"}{"},%
    morestring=[s]{u'''}{'''},%
    morestring=[s]{u"""}{"""},%
    identifierstyle=\color{black}\ttfamily,
    commentstyle=\color{mathematica_cyan}\ttfamily,
    stringstyle=\color{mathematica_red}\ttfamily,
    keepspaces=true,
    showspaces=false,
    showstringspaces=false,
    rulecolor=\color{mathematica_frame},
    frame=single,
    frameround={t}{t}{t}{t},
    framexleftmargin=10mm,
    numbers=left,
    numberstyle=\tiny\color{halfgray},
    backgroundcolor=\color{mathematica_bg},
    basicstyle=\footnotesize,
    keywordstyle=\color{mathematica_green}\ttfamily,
    aboveskip=1.2em,
    belowskip=1.2em,
}

\let\origthelstnumber\thelstnumber
\makeatletter
\newcommand*\Suppressnumber{%
  \lst@AddToHook{OnNewLine}{%
    \let\thelstnumber\relax%
     \advance\c@lstnumber-\@ne\relax%
    }%
}

\newcommand*\Reactivatenumber{%
  \lst@AddToHook{OnNewLine}{%
   \let\thelstnumber\origthelstnumber%
   \advance\c@lstnumber\@ne\relax}%
}

\def\lagr{\mathcal{L}}

\def\thv[#1,#2,#3]{\left( \begin{array}{c} #1 \\ #2 \\ #3 \end{array} \right)}
\def\twv[#1,#2]{\left( \begin{array}{c} #1 \\ #2 \end{array} \right)}

\def\ov{\overline}

\def\nn{\nonumber}
\def\bra{\langle}
\def\ket{\rangle}

\def\beq{\begin{equation}}
\def\eeq{\end{equation}}

\newcommand\SARAH{{\tt SARAH}\xspace}

\newcommand\SPheno{{\tt SPheno}\xspace}

\newcommand\Mathematica{{\tt Mathematica}\xspace}

\newcommand{\newc}{\newcommand}

\newc{\Psibar}{\overline{\Psi}}
\newc{\FFbS}{\overline{FF}S}
\newc{\FFbV}{\overline{FF}V}
\newc{\FFbe}{\overline{FF}\epsilon}
\newc{\FSS}{F_{SS}}
\newc{\FSSS}{F_{SSS}}
\newc{\FFFS}{F_{FFS}}
\newc{\FFFbS}{F_{\overline{FF}S}}
\newc{\FSSV}{F_{SSV}}
\newc{\FVS}{F_{VS}}
\newc{\FVVS}{F_{VVS}}
\newc{\FFFV}{F_{FFV}}
\newc{\FFFbV}{F_{\overline{FF}V}}
\newc{\FVV}{F_{VV}}
\newc{\FVVV}{F_{VVV}}
\newc{\fggV}{f_{ggV}}
\newc{\Fgauge}{F_{\rm gauge}}
\newc{\fSS}{f_{SS}}
\newc{\fSSS}{f_{SSS}}
\newc{\fFFS}{f_{FFS}}
\newc{\fFFbS}{f_{\overline{FF}S}}
\newc{\fSSV}{f_{SSV}}
\newc{\fVVS}{f_{VVS}}
\newc{\fVS}{f_{VS}}
\newc{\fFFV}{f_{FFV}}
\newc{\fFFbV}{f_{\overline{FF}V}}
\newc{\fVV}{f_{VV}}
\newc{\fVVV}{f_{VVV}}
\newc{\fgauge}{f_{\rm gauge}}
\newc{\FFVbar}{{\overline{FF}V}}
\newc{\FFSbar}{\overline{FF}S}

\newc{\bsigmu}{\bar\sigma^\mu}
\newc{\DRbar}{\overline{\text{DR}}}
\newc{\DRbarprime}{\overline{\text{DR}}^\prime}

\begin{document}

\title{Unitarity constraints on general scalar couplings with \SARAH}
\author{
   Mark D. Goodsell \thanksref{a1} \and
   Florian Staub \thanksref{a2,a3}
   }

\institute{ Laboratoire de Physique Th\'eorique et Hautes Energies (LPTHE),\\ UMR 7589,
Sorbonne Universit\'e et CNRS, 4 place Jussieu, 75252 Paris Cedex 05, France\label{a1}
\and
Institute for Theoretical Physics (ITP), Karlsruhe Institute of Technology, Engesserstra{\ss}e 7, D-76128 Karlsruhe, Germany \label{a2}
\and
Institute for Nuclear Physics (IKP), Karlsruhe Institute of Technology, Hermann-von-Helmholtz-Platz 1, D-76344 Eggenstein-Leopoldshafen, Germany \label{a3}
}

\date{}

\maketitle

\begin{abstract}
We present an update of the \Mathematica package \SARAH to calculate unitarity constraints in BSM models. The new functions can perform an analytical and 
numerical calculation of the two-particle scattering matrix of (uncoloured) scalars. We do not make use of the simplifying assumption of a very large scattering energy, but include all contributions which could become important at small energies above the weak scale. This allows us to constrain trilinear scalar couplings. However, it can also modify (weakening or strengthening) the constraints on quartic couplings, which we show via the example of a singlet extended Standard Model. 
\end{abstract}

\section{Introduction}
In a classic paper, Lee, Quigg and Thacker showed that the Higgs mass in the Standard Model (SM) must be below 1~TeV in order to maintain perturbative unitarity \cite{Lee:1977eg}. From the measurement of the Higgs mass at the LHC \cite{Aad:2012tfa,Chatrchyan:2012xdj}, we have learned that the quartic coupling in the SM is even 
well below 1, i.e. the scalar sector of the SM has very weak self-couplings. However, this is not necessarily true if one adds more fundamental scalars to the theory. The scalar potential
of BSM models often involve many new parameters which are experimentally barely constrained. Therefore, theoretical conditions like the stability of the potential or
the conservation of unitarity are very important to find physical viable parameter regions in these models. 

The constraints from tree-level perturbativity are often applied in well studied models such as ones with additional singlets \cite{Cynolter:2004cq,Kang:2013zba,Costa:2014qga}, 
doublets \cite{Casalbuoni:1986hy,Casalbuoni:1987eg,Maalampi:1991fb,Kanemura:1993hm,Ginzburg:2003fe,Akeroyd:2000wc,Horejsi:2005da}, or triplets \cite{Khan:2016sxm,Aoki:2007ah,Hartling:2014zca}.
However, very often the constraints are derived under the assumption that the scattering energy is much larger than the involved masses. In this limit, only point interactions 
are important, and all diagrams with propagators are neglected. As a consequence, cubic couplings do not enter the widely used constraints at all. In this work, we shall present a general calculation of unitarity constraints without this assumption, with the following motivation:
\begin{enumerate}
\item We want to place bounds on genuine trilinear couplings.
\item For theories where additional scalars couple to the Higgs, even if there are no trilinear couplings before electroweak symmetry breaking, they are generated after the Higgs takes a vev, and unitarity of scattering at finite $s$ gives new constraints on these \emph{quartic} couplings.
\item For theories defined with a low cutoff, scattering may never be in the regime where the energies are sufficiently large to neglect the $s,t,u$--channel processes.
\item Even for theories with a high cutoff, the infinite energy approximation is rarely justified since the couplings must run: if we take the energy sufficiently high to be able to neglect particle masses, the resummed couplings will typically have completely different values. 
\end{enumerate}

For this purpose we have extended the 
\Mathematica package \SARAH \cite{Staub:2008uz,Staub:2009bi,Staub:2010jh,Staub:2012pb,Staub:2013tta} by routines for the analytical and numerical study of the full tree-level unitarity constraints. While the analytical routines are helpful to 
obtain expressions for $2\to 2$ scattering elements, a symbolic calculation of the full scalar scattering matrix could become slow and less illuminating. Therefore, for practical application
the Fortran output for \SPheno \cite{Porod:2003um,Porod:2011nf} has been also extended to obtain a numerical prediction for the maximal eigenvalue of the full scattering matrix.

We discuss in sec.~\ref{sec:generic} the underlying calculations to obtain unitarity constraints in generic BSM models, and the assumptions/restrictions that we shall apply. The importance 
of the full calculation is demonstrated in sec.~\ref{sec:example} via the example of singlet extensions of the SM. In sec.~\ref{sec:implementation} we show how the new routines are used. A brief summary is given in sec.~\ref{sec:summary}.

\section{Generic calculation of unitarity constraints}
\label{sec:generic}
\subsection{$2\to 2$ Scattering processes of scalars at finite momentum}
The derivation of unitarity constraints is elementary, but the derivation for finite momentum is rarely found in the literature -- and there are many common misunderstandings -- so we present a clear exposition in appendix \ref{APP:partialwaves}. The result is that the partial wave constraint becomes
\begin{align}
-i (a_J - a_J^\dagger) \le a_J a_J^\dagger \qquad \forall J 
\label{EQ:BasicUnitarity}\end{align}
where $a_J$ is a normal matrix related to the partial wave decomposition of $2 \rightarrow 2$ scattering matrix elements $\mathcal{M}_{ba}$ from a scattering of a pair of particles $a= \{1,2\}$ with momenta $\{p_1, p_2\}$ to a pair $b = \{3,4\}$ with momenta $\{k_3, k_4\}$ as
\begin{align}
a_J^{ba} \equiv& \frac{1}{32\pi} \sqrt{\frac{4 |\mathbf{p}^b| |\mathbf{p}^a|}{2^{\delta_{12}} 2^{\delta_{34}}\, s}} \int_{-1}^1 d(\cos \theta) \mathcal{M}_{ba} (\cos \theta) P_J (\cos \theta).
\end{align}
The factor  $\delta_{12} (\delta_{34})$ is $1$ if particles $\{1,2\} (\{3,4\})$ are identical, and zero otherwise. $P_J$ are the Legendre polynomials, $\mathbf{p}^i$ is the centre of mass three-momentum for particle $i$, and $s=(p_1 + p_2)^2$ is the standard Mandelstam variable.

From the fundamental equation (\ref{EQ:BasicUnitarity}) different constraints can be derived; we shall only consider the zeroth partial wave, and denoting $a_0^i$ as the eigenvalues of $a_0$ we shall apply 
\begin{align}
\mathrm{Re} (a_0^i) \le& \frac{1}{2}\  \forall\  i.
\end{align}

\begin{figure}[tb]
\centering
\includegraphics[width=0.66\linewidth]{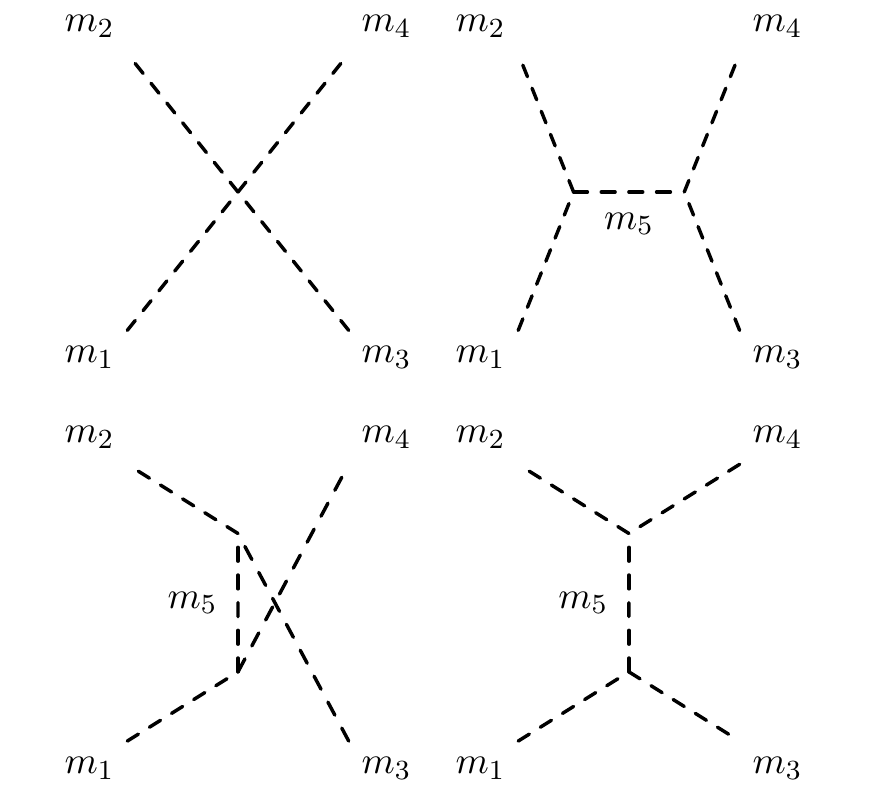} 
\caption{The four topologies contributing to $2 \to 2$ scalar scattering processes at finite $\sqrt{s}$. In the approximation of $\sqrt{s} \gg m_i$, only the point interaction contributes.}
\label{fig:diagrams}
\end{figure}

The diagrams which contribute to $2 \to 2$ scalar scattering processes are shown in Fig.~\ref{fig:diagrams}. For a general field theory consisting of real scalars $\phi_i$ and couplings
\begin{align}
  \mathcal{L} \supset - \frac{1}{6} \kappa^{ijk} \phi_i \phi_j \phi_k - \frac{1}{24} \lambda^{ijkl} \phi_i \phi_j \phi_k \phi_l
\end{align}
the matrix elements are
\begin{align}
\mathcal{M} (1,2 \rightarrow 3,4) =&  -  \lambda^{1234} -\kappa^{12 5} \kappa^{345} \frac{1}{s - m_5^2} -\kappa^{13 5} \kappa^{245} \frac{1}{t - m_5^2} - \kappa^{14 5} \kappa^{235} \frac{1}{u - m_5^2}.
\end{align}

The integration over $\cos\theta$ is trivial for the contact and $s$-channel processes, and always straightforward for the others using
\begin{align}
t =& m_1^2 + m_3^2 - 2 E_1 E_3 + 2 |\mathbf{p}_1| |\mathbf{p}_3| \cos \theta \qquad u = m_1^2 + m_4^2 - 2 E_1 E_4 - 2 |\mathbf{p}_1| |\mathbf{p}_3| \cos \theta,
\end{align}
where $E_i$ are the energies of the particles in the centre of mass frame, and $\mathbf{p}_1, \mathbf{p}_3$ are the three-momenta. We shall express the results in terms of the function
\begin{align}
\lambda(s, m_i^2, m_j^2) \equiv& \frac{1}{s^2} \bigg[s^2 + m_i^4 + m_j^4 - 2 m_i^2 m_j^2 - 2 s m_i^2 - 2 s m_j^2\bigg],
\end{align}
so that
\begin{align}
  |\mathbf{p}_1| =& \frac{1}{2} \sqrt{s\lambda(s, m_1^2, m_2^2)}, \qquad |\mathbf{p}_3| = |\mathbf{p}_4|=\frac{1}{2} \sqrt{s\lambda(s, m_3^2, m_4^2)} \nn\\
  E_1 =& \frac{s + m_1^2 - m_2^2}{2 \sqrt{s}}, \qquad E_3 = \frac{s + m_3^2 - m_4^2}{2 \sqrt{s}}
\end{align}
allowing us to define
\begin{align}
  f_t (s, m_1^2, m_2^2, m_3^2, m_4^2,m_5^2) \equiv& \frac{1}{2} \sqrt{\frac{4 |\mathbf{p}_1| |\mathbf{p}_3|}{s}} \int_{-1}^1 \text{d}\cos\theta \frac{1}{t-m_5^2} \nn\\
  =& \frac{1}{s} \frac{1}{[\lambda(s, m_1^2, m_2^2) \lambda(s, m_3^2, m_4^2)]^{1/4}} \log \left( \frac{m_1^2 + m_3^2 - m_5^2 - 2 E_1 E_3 + 2 |\mathbf{p}_1| |\mathbf{p}_3|}{m_1^2 + m_3^2 - m_5^2 - 2 E_1 E_3 -2 |\mathbf{p}_1| |\mathbf{p}_3|}\right)\nn \\
  f_u (s, m_1^2, m_2^2, m_3^2, m_4^2,m_5^2) \equiv& f_t(s, m_1^2, m_2^2, m_4^2, m_3^2,m_5^2).
  \end{align}
  In terms of these, the (modified) zeroth partial waves are
  \begin{align}
    a_0 =&- \frac{2^{-\frac{1}{2}(\delta_{12}+ \delta_{34})}}{16 \pi } \bigg\{ \bigg[\lambda(s, m_1^2, m_2^2) \lambda(s, m_3^2, m_4^2)\bigg]^{1/4} \bigg[  \lambda^{1234} + \kappa^{12 5} \kappa^{345} \frac{1}{s - m_5^2}\bigg] \nn\\
         & - \kappa^{13 5} \kappa^{245} f_t (s, m_1^2,m_2^2, m_3^2, m_4^2,m_5^2) - \kappa^{14 5} \kappa^{235} f_u (s, m_1^2,m_2^2, m_3^2, m_4^2,m_5^2)\bigg\}.
  \end{align}

\subsection{Handling of poles}
In the neighbourhood of poles, the tree-level amplitude diverges which signals that we need to take higher-order corrections into account (which will effectively modify the divergent propagator to include the width, cutting off the divergence). Moreover, since we are only calculating unitarity constraints at tree-level, in the presence of large couplings large quantum corrections to masses may mean that the physical location of the poles is a long way away from the tree-level mass parameters. Both of these issues imply that we should not trust our results in such cases. We therefore apply the following conditions:
\begin{enumerate}
\item {\bf $s$-Channel poles}
Obviously, $s$-channel poles are present if any propagator mass is close to $\sqrt{s}$. In order to cut out this region, we set the {\it entire irreducible scattering matrix} to zero if the condition
\begin{equation}
|1 - \frac{s}{m^2}| > 0.25 
\end{equation}
is violated.
\item {\bf $t$-/$u$-Channel poles}
Particles in the $t$ and $u$ channels can become on-shell. For a $t$-channel diagram, this can happen if
\begin{equation}
m_1 > m_3 + m_5 \hskip1cm \vee \hskip1cm m_3 > m_2 + m_5 
\end{equation}
holds. Similar conditions exist, for $1 \leftrightarrow 2$ and $3 \leftrightarrow 4$. The conditions for $u$-channels are obtained by exchanging $3 \leftrightarrow 4$. These conditions (used in \cite{Schuessler:2007av}) 
are only necessary to have a pole, but not sufficient -- they are too conservative. In fact, the presence of such a pole  also demands that the scattering energy $s$ is smaller than a given value. The general conditions for 
the minimal scattering energy $s_{\rm min}$ to avoid poles are
\begin{align}
s_{\rm min,t} = & \frac{1}{2 m_5} \Big(\sqrt{m_1^2-2 m_1 (m_3+m_5)+(m_3-m_5)^2} \sqrt{m_2^2-2 m_2 (m_4+m_5)+(m_4-m_5)^2}+ \nonumber \\
 & \hskip2cm m_1 (-m_2+m_4+m_5)+m_2 m_3+m_2 m_5-m_3 m_4+m_3 m_5+m_4 m_5-m_5^2 \Big)  \\
s_{\rm min,u} = & \frac{1}{2 m_5} \Big(\sqrt{m_1^2-2 m_1 (m_4+m_5)+(m_4-m_5)^2} \sqrt{m_2^2-2 m_2 (m_3+m_5)+(m_3-m_5)^2}+ \nonumber \\
 & \hskip2cm m_1 (-m_2+m_3+m_5)+m_2
   m_4+m_2 m_5-m_3 m_4+m_3 m_5+m_4 m_5-m_5^2 \Big)   
\end{align}
From this we find that for an often appearing, kinematic configuration 
with $m_3=m_1$ and $m_4 = m_2$, a $t$-channel pole only shows up for 
\begin{equation}
s < m_1 + m_2 + \frac12 \left(-m_5 + \sqrt{(-4 m_1 + m_5)(-4 m_2 + m_5)}\right)
\end{equation}
We will include three different treatments of such poles, which the user can select depending on taste:
\begin{enumerate}
 \item Only the matrix element for which such a poles appears is set to zero, but all other entries of the scattering matrix are kept. This gives the most aggressive limits. 
 \item A partial diagonalisation of the scattering matrix is performed as proposed in Ref.~\cite{Schuessler:2007av}
 \item The entire irreducible scattering matrix is set to zero. This gives the weakest limits.
\end{enumerate}
\end{enumerate}

\subsection{The role of the Goldstone boson equivalence theorem}
\label{sec:GBE}

To apply unitarity constraints, in principle we should consider all coupled channels for all particles. However, in practice there are a large number available, most of which will not contribute in a meaningful way to constraints, and so in the interest of computational speed it is necessary to impose some simplifying assumptions. These are:
\begin{enumerate}
\item We can neglect all contributions proportional to gauge couplings, and scatter at energies well  above the mass of any gauge bosons; clearly for smaller scattering energies this would mean we would be in the neighbourhood of an abundance of poles. Furthermore, light bosons mediate infra-red unsafe scattering, so our above formalism would require modification, and it is therefore reasonable to eliminate them.
\item We neglect all fermionic contributions. The above assumption partly justifies this, as the contributions to scattering from Standard Model fermions should be small at energies well above their masses.
\item To avoid an abundance of group structures of the scattering pairs, we do not consider any particles that transform under any unbroken symmetries except for the electric charge. In particular, this excludes any strongly coupled particles (such as top partners).
\end{enumerate}
Assumption (1) is the most reasonable, and also most powerful: since amplitudes involving \emph{transverse} gauge bosons are always proportional to gauge couplings, we can neglect them. For longitudinal gauge bosons of mass $m_V$, whose polarisation vectors can be taken to be
$$
\epsilon^\mu = \frac{1}{m_V} ( |\mathbf{p}|, E \frac{\mathbf{p}}{|\mathbf{p}|}),
$$
the scattering amplitudes contain factors of $1/m_V$ and hence inverse powers of the gauge couplings, so that they can have a finite amplitude as the gauge couplings are taken to zero. On the other hand, since we scatter at energies well above their masses, the Goldstone Boson equivalence theorem allows us to instead replace all \emph{external} longitudinal gauge bosons with the Goldstone boson, with the important proviso that it has a physical mass equal to the gauge boson mass (so not equal to $\xi m_V$). It then turns out that, since we neglect contributions proportional to the gauge couplings, we can also neglect gauge boson \emph{propagators} -- but only if we work in Feynman gauge; we discuss in appendix  \ref{APP:GaugeProp} why this is so and what happens in other gauges. 

Taken together, then, the above assumptions, and working in Feynman gauge, enable us to consider scattering amplitudes where all states are scalars. While it would be an interesting if time-consuming task to relax some of these assumptions (which we leave to future work), they are already very powerful and allow us to study a wide range of theories.

\section{Examples: singlet extentions of the Standard Model}
\label{sec:example}
We want to demonstrate the importance of the unitarity constraints beyond the large $s$ approximation by a brief example.

\subsection{Pure singlet model}

First we shall consider the simplest possible BSM model: the SM extended by a real singlet $S$. To illustrate point (1) in the introduction, if we just consider the singlet and assume that its couplings to the Higgs sector are small relative to its self-couplings, we can take the Lagrangian:
\begin{align} \mathcal{L} \supset& -\frac{1}{2} m_S^2 S^2 - \frac{1}{3} \kappa S^3 - \frac{1}{2}\lambda_S S^4 .
 \label{EQ:PureSinglet} \end{align}
 There are two additional minima away from the origin if
  $$ \kappa^2 > 8 m_S^2 \lambda_S, $$
  but the origin remains the true minimum if     
  \begin{align}
  \kappa^2 < 12 m_S^2 \lambda_S \pm \kappa \sqrt{ \kappa^2 - 8 m_S^2 \lambda_S} \hspace{1cm} \longrightarrow \hspace{1cm}  |\kappa/m_S| < 3 \sqrt{\lambda_S}.
  \end{align}
As a probe of genuine trilinear couplings, taking the minimum at the origin is most interesting, because once the singlet obtains an expectation value, stability constraints the trilinears to be rather small compared to the physical mass.

\begin{figure}\begin{center}
    \includegraphics[width=0.5\textwidth]{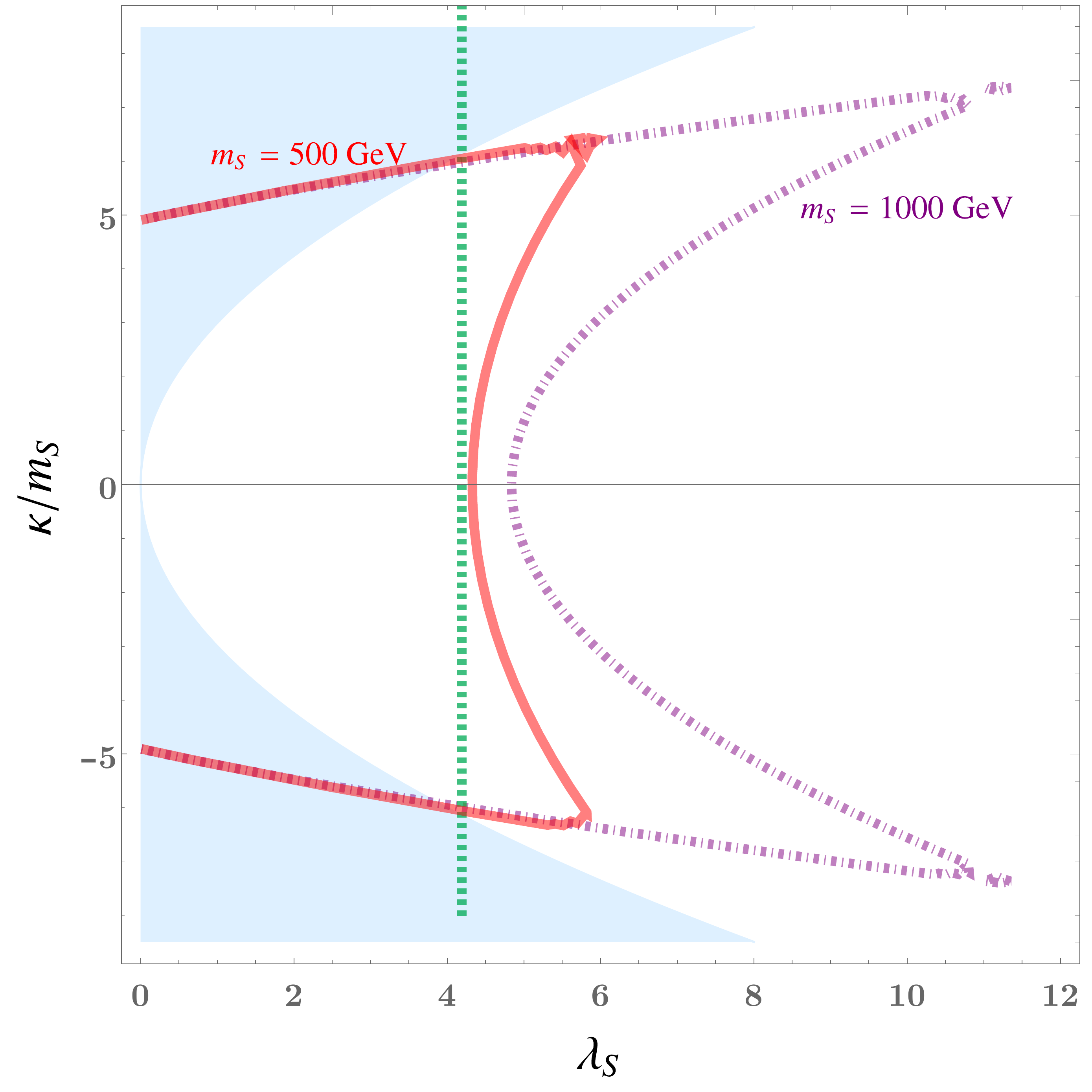}\includegraphics[width=0.5\textwidth]{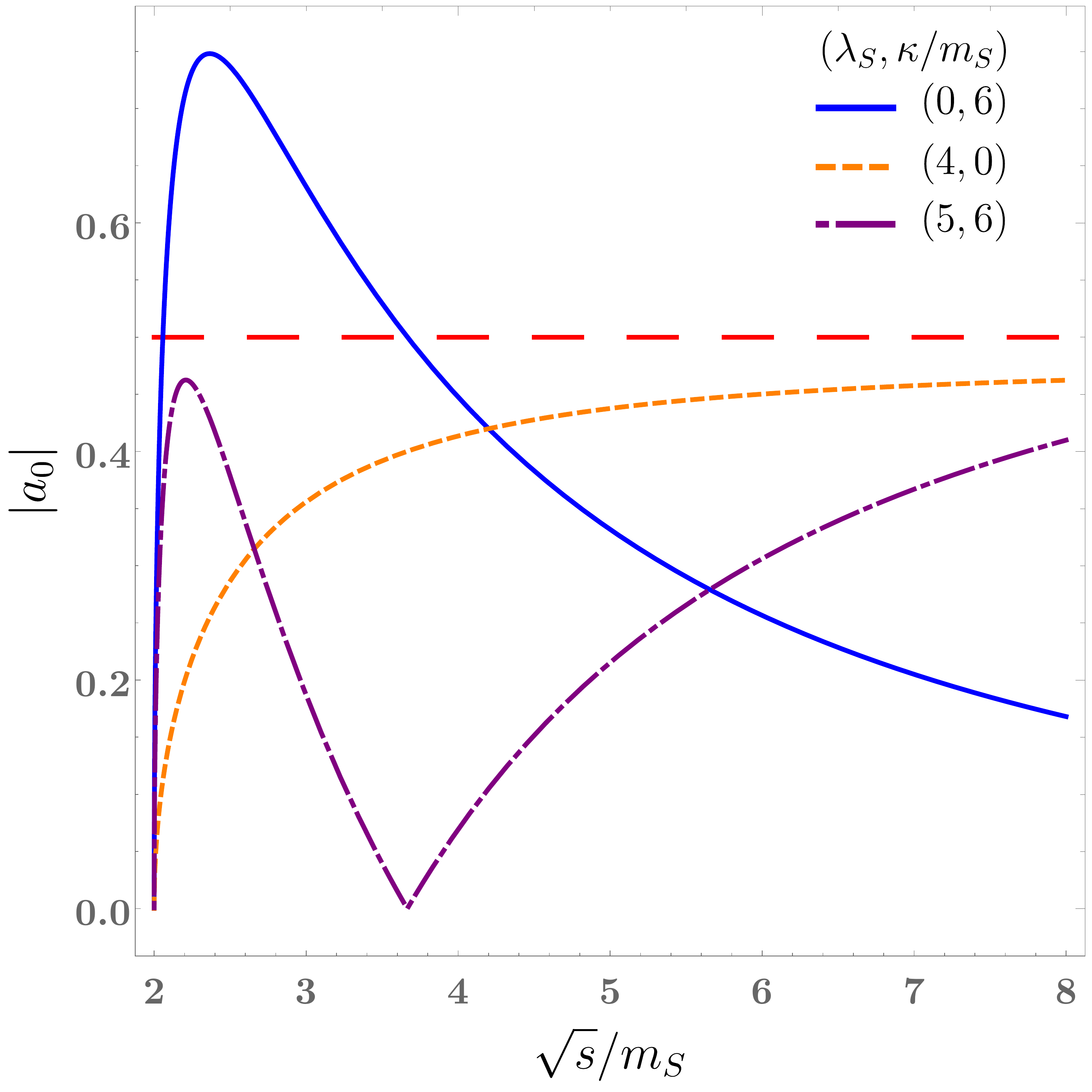}
    \caption{Unitarity constraints on the pure singlet model (\ref{EQ:PureSinglet}). \emph{Left}: the solid red and dashed purple lines correspond to $m_S = 500, 1000$ GeV respectively (or other arbitrary units relative to $\sqrt{s} \le 4000$ GeV). The blue shaded regions are excluded by stability of the vacuum. \emph{Right}: $|a_0|$ vs $\sqrt{s}/m_S$ for different values of the couplings $\lambda_S, \kappa/m_S$.}
\label{FIG:trilinear}\end{center}
\end{figure}

It is simple to derive $a_0$ for this case:
  \begin{align}
  a_0 =& - \frac{1}{32\pi} \bigg[\sqrt{1- \frac{4m_S^2}{s}} \bigg( 12 \lambda_S + \frac{4 \kappa^2}{s-m_S^2}\bigg)   + \frac{8 \kappa^2}{\sqrt{s(s -4 m_S^2)}} \log \frac{ m_S^2}{s - 3m_S^2} \bigg].
\end{align}
We show the constraints on this for $\sqrt{s} = 4000$ GeV, $m_S = 500$ and $1000$ GeV (or other arbitrary units) in figure \ref{FIG:trilinear}. Clearly the constraint from $s \rightarrow \infty$ would give $\lambda_S < \frac{4\pi}{3},$ and this is shown as the green vertical dashed line of the left-hand plot. To understand the role of the scattering energy, we show the behaviour of $|a_0|$ as $\sqrt{s}$ is varied above threshold on the right-hand plot: there is always a rapid increase followed by logarithmic behaviour. 

The finite $s$ constraints clearly consist of two different regimes that intersect, and come from the fact that the $t/u$ channel contribution has opposite sign to the $s$-channel and quartic term. As a function of $s$, $a_0$ grows sharply from $0$ at $s = 4 m_S^2$, before decreasing again and tending to the large $s$ value. So the constraints come both from the maximum allowed $s$, and  around $s = 6 m_S^2$; for $\lambda_S = 0$ it occurs at $s \simeq 5.6 m_S^2$. From the maximum $s$, we obtain the curved regions in the plot that have a minimum value for $\lambda_S$ as it passes through $\kappa =0$. These therefore show a difference when we change $m_S$. On the other hand, the overlapping curves that pass into the unstable region $|\kappa/m_S| > 3 \sqrt{\lambda_S}$ come from taking $s$ near $6 m_S^2$. Note that the value $ s = 5.6 m_S^2$ is not near any pole value, and the corrections to the singlet mass are well under control; at one loop they are
\begin{align}
\delta m_S^2 =\frac{2 \kappa^2}{16\pi^2} B_0 (p^2, m_S^2, m_S^2) + \frac{6\lambda_S}{16\pi^2} A_0 (m_S^2),
\end{align}
so for $\kappa = 5 m_S$, we have $\delta m_S^2 \sim 0.3 m_S^2$. Moreover, the scattering energy is sufficiently large that the produced particles are relativistic, so we are not in a regime where e.g. Sommerfeld enhancements would play a significant role. Hence the enhancement to the partial wave amplitude is a genuine physical effect that we can use to constrain the couplings of the theory. In particular, it gives an upper bound on $\kappa$, independently from vacuum stability considerations -- especially for larger values of $\lambda_S$. We expect this to be a general feature: from an inspection of the right-hand plot of figure \ref{FIG:trilinear} we see that the strongest limits (away from poles) to a given model will either come from near-threshold production or at large $s$.  

This model also allows us to simply illustrate our points (3) and (4) in the introduction. What we are interested in constraining are the values of $\kappa/m_S$ and $\lambda_S$ at low energies. However, the partial waves receive quantum corrections which can be very significant for large couplings, and if the scattering energy is large, we should certainly resum the logarithms and place constraints only on couplings evaluated \emph{at a renormalisation scale of $\sqrt{s}$} (see e.g. \cite{Grinstein:2015rtl}). In this model, the one-loop $\beta$-function for the quartic coupling gives
\begin{align}
\frac{d \lambda_S}{d \log \mu} =& \frac{36 \lambda_S^2}{16 \pi^2},
\end{align}
which can be solved exactly, and gives a Landau pole at
\begin{align}
\mu =& m_S \exp \left[ \frac{4\pi^2}{9 \lambda_S (m_S)} \right].
\end{align}
For $\lambda_S (500\ \mathrm{GeV}) = 4,$ this is at $1500$ GeV! Hence we cannot apply the infinite-energy scattering limit to this coupling. Put another way, since we must understand the limits in figure \ref{FIG:trilinear} to be evaluated at $\mu = \sqrt{s}$, if $\lambda_S^{\rm max} (4000\ \mathrm{GeV}) = 4$, then $ \lambda_S^{\rm max} (500\ \mathrm{GeV}) = 1.4.$

\subsection{Singlet extended SM with conserved $Z_2$}

While the above model is trivial, it contains most of the ingredients that we find in more complicated models, in particular the partial cancellation between the channels. Now we will turn to a more physical example: a singlet that couples to the Higgs, but with a $Z_2$ symmetry which stabilises it and prevents mixing with the Higgs. The potential reads
\begin{equation}
V = \frac12 \lambda_H |H|^4   + \frac12 \lambda_{HS} |H|^2 S^2 + \frac12 \lambda_S S^4 + m_H^2 |H|^2 + \frac12 m_S^2 S^2
\end{equation}
This theory contains no trilinear scalar couplings before electroweak symmetry breaking. As such, it is a useful prototype of popular extensions of the SM such as the Two Higgs Doublet Model, NMSSM, etc, as well as being phenomenologically interesting in its own right (for example, it provides a dark matter candidate). However, once the Higgs obains an expectation value so that we can write the neutral Higgs boson $H^0 = \frac{1}{\sqrt{2}} (v + h + i G)$, a trilinear coupling
$$ \lagr \supset - \frac{1}{2} v \lambda_{HS} h S^2$$
is generated.
Thus we will have $s, t, u$-channel scattering processes in the scalar sector which will modify the unitarity constraints!

In the large $s$ limit we have
\begin{align}
\text{Max}\left\{\left|\lambda_{HS}\right| ,\left| \lambda_H \right| ,\frac{1}{2} \left|6 \lambda_{S}+3 \lambda_H\pm\sqrt{4 \lambda_{HS}^2+36 \lambda_{S}^2+9 \lambda_H^2-36 \lambda_{S} \lambda_H }\right| \right\}<8 \pi
\end{align}
We want to compare this with the full calculation. Results for the scattering processes are already given in literature \cite{Cynolter:2004cq,Kang:2013zba}, but we disagree with both references in different channels. Therefore, we 
list all matrix elements in appendix~\ref{app:singlet}. In the following, analytical discussion, we concentrate only on the parts involving CP-even states. 
The scattering matrix involving only the the Higgs and the singlet is 
\begin{equation}
\begin{pmatrix}
hh\to hh & hh \to SS & 0 \\
SS \to hh & SS \to SS & 0 \\
0 & 0 & hS\to hS
\end{pmatrix}
\end{equation}
If we assume for the moment $\lambda_{HS} \gg \lambda, \lambda_S$, the dominant contribution is the $hS\to hS$ scattering. 

\begin{figure}[tb]
\centering
\includegraphics[width=0.5\linewidth]{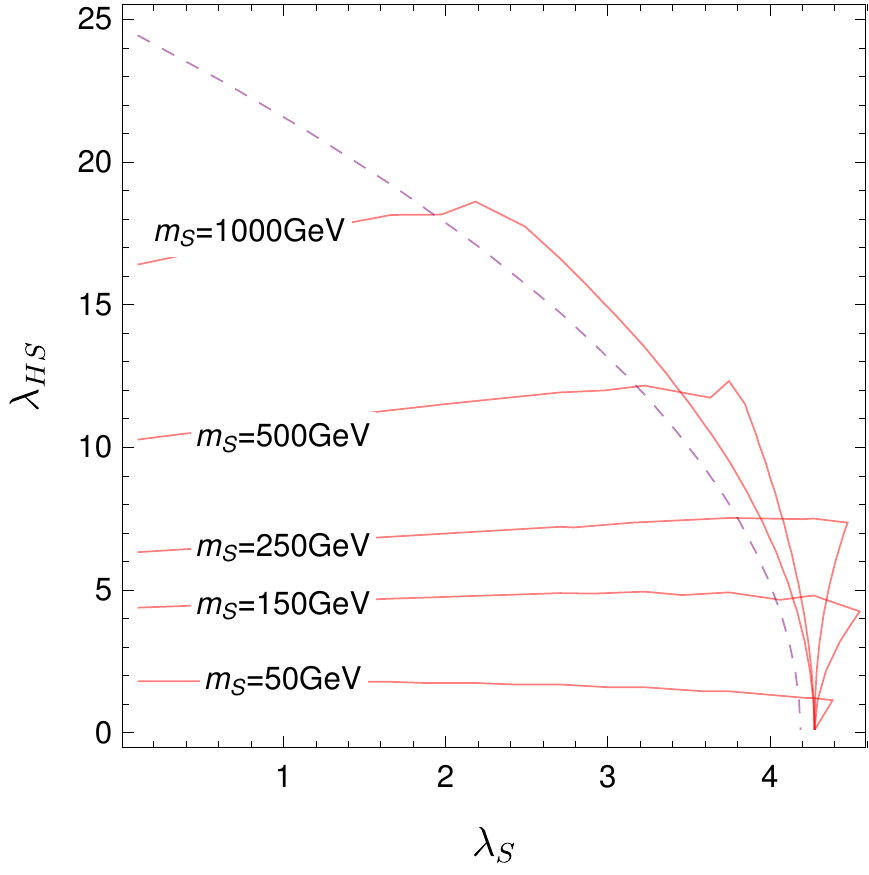}
\caption{Unitarity constraints for the singlet extended SM. The dashed purple line gives the limit when using only point interactions in the large $s$ limit. The red lines give the constraints for the full calculation for different singlet masses.}
\label{fig:SSMZ2}
\end{figure}

The
result reads
\begin{align}
16\pi a_0(hS\to hS)=&-\frac{\lambda_{HS}}{16 \pi  s \left(s-m_S^2\right) \sqrt{m_h^4-2 m_h^2
   \left(m_S^2+s\right)+\left(m_S^2-s\right)^2}} \times \nonumber \\
   & \Bigg[-\left(m_h^4-2 m_h^2 \left(m_S^2+s\right)+\left(m_S^2-s\right)^2\right) \left(-\lambda_{HS} v^2+m_S^2-s\right) \nonumber \\
   & +\lambda_{HS} s v^2 \left(s-m_S^2\right)
   \log \left(\frac{m_h^4-2 m_h^2 m_S^2+m_S^4-m_S^2 s}{s \left(2 m_h^2+m_S^2-s\right)}\right) \nonumber \\
   & +3 m_h^2 s \left(s-m_S^2\right) \log \left(\frac{m_h^2
   s}{m_h^4-m_h^2 \left(2 m_S^2+s\right)+\left(m_S^2-s\right)^2}\right)\Bigg]
\end{align}
In order to simplify this expression we consider the limit of small $m_S$ and large $v^2\lambda_{HS}^2 \gg m_h^2 \gg m_S^2$. This results in
\begin{align}
16\pi a_0(hS\to hS)\simeq&-\frac{\lambda_{HS}^2 v^2}{s^2 \left(s-m_h^2\right)} \left(\left(m_h^2-s\right)^2+s^2 \log \left(\frac{m_h^4-2 m_h^2 m_S^2-m_S^2 s}{2 m_h^2 s-s^2}\right)\right)
\end{align}
Thus, for $s \sim m_h^2$, this scales as
\begin{equation}
16\pi a_0(hS\to hS)\sim \frac{\lambda_{HS}^2 v^2}{m_h^2}
\end{equation}
which can be significantly larger than the limit from point interactions only. This is also confirmed by our numerical calculation with \SPheno. In Fig.~\ref{fig:SSMZ2} we compare the limits from including point interaction only with the full calculation in the $(\lambda_S,\lambda_{HS})$ plane for different singlet masses. 
We see that the unitarity limits on $\lambda_{HS}$ become much stronger for $m_S < m_h$ and small $\lambda_S$. However, even for larger masses a pronounced effect is visible. Even for $m_S=500$~GeV, the limits are stronger by a factor of two. \\

This brief example demonstrates the importance of going beyond the large $s$ approximation when considering unitarity constraints in BSM models. Detailed discussions of these effects in other, phenomenologically more interesting models will be given elsewhere \cite{inprepthdm,inpreptriplet}.

\section{Implementation in \SARAH}
\label{sec:implementation}
We have extended the \Mathematica package by the results and procedures summarised in the previous sections; in particular, the restrictions/assumptions that we apply are described in section \ref{sec:GBE}. The user has two possibilities to use the new functionality: (i) during the 
\Mathematica session analytical expressions for specific scattering processes or the full scattering matrix are available; (ii) the necessary routines for a numerical calculation of the 
unitarity constraints are included in the \SPheno output. Below, we give some details how to work with both methods.

\subsection{Interactive Session in {\tt Mathematica}}
\subsubsection{Commands}
In order to obtain analytical expressions for $2\to 2$ scattering processes or the entire scattering matrix new commands are available in \SARAH with version 4.13.0. 
\begin{enumerate}
\item {\bf Initialisation:} in order to initialise the calculation of the unitarity constraints, one needs to run
\begin{MIN}
InitUnitarity[{assumptions}] 
\end{MIN}

This command calculates all necessary (scalar) vertices. In addition, a list with assumptions can be given which is used to modify the appearance of the vertices. Possible assumptions are:
\begin{itemize}
 \item Some parameters are neglected, e.g. 
\begin{MIN}
InitUnitarity[{LambdaS->0,LambdaH->0}] 
\end{MIN}
\item Mixing between scalars are neglected by replacing rotation matrices with a Kronecker Delta, e.g. 
\begin{MIN}
InitUnitarity[{ZH->Delta}] 
\end{MIN}
\item Some couplings are expressed in terms of other parameters, e.g.
\begin{MIN}
InitUnitarity[{Lambda->mh^2/v^2}] 
\end{MIN}
\end{itemize}
\item {\bf Scattering processes}: once the vertices are initialised, specific scattering processes are obtained via
\begin{MIN}
GetScatteringDiagrams[{incoming1, incoming2} -> {outgoing1, outgoing2}] 
\end{MIN}
Here, {\tt incoming1, incoming2} are the incoming particles and {\tt outgoing1, outgoing2} the outgoing ones. One needs to use for these variables the names of fields in \SARAH. Optionally, a generation index can also be given. 
\begin{itemize}
 \item No explicit generation indices, e.g. 
 \begin{MIN}
GetScatteringDiagrams[{hh,hh} -> {Ah, Ah}] 
\end{MIN}
This returns the scattering element for $hh\to AA$. If the scalar $h$ and pseudo-scalar $Ah$ appears in several generations in the given model, the indices {\tt in1}, {\tt in2}, {\tt out1}, {\tt out2} are used.  
\item Explicit generation indices, e.g.
\begin{MIN}
GetScatteringDiagrams[{hh[1],hh[1]} -> {Ah[2], Ah[2]}] 
\end{MIN}
This sets the generation indices of the incoming fields to 1 and of the outgoing fields to 2. 
\end{itemize}
The result of {\tt GetScatteringDiagrams} is a function of the couplings and masses in the model. In addition, 
keywords are introduced to make it possible to trace back the origin of the different terms: $s$, $t$ and $u$-channel diagrams as well as point interactions are multiplied with a variable:
\begin{itemize}
\item {\tt sChan} for a $s$-channel diagram
\item {\tt tChan} for a $t$-channel diagram
\item {\tt uChan} for a $u$-channel diagram
 \item {\tt qChan} for quartic interactions
\end{itemize}
Thus, one can easily remove specific diagrams in order to check their impact by setting the corresponding variables to zero. 
\item {\bf Scattering matrix}: the full scattering matrix is return by running
\begin{MIN}
BuildScatteringMatrix 
\end{MIN}
All generation indices of the external fields are explicitly inserted. 
\end{enumerate}

\subsubsection{Example}
We show via the example of the SM how the new commands are used in practice. First, \SARAH needs to be loaded and the SM be initialised:
\begin{MIN}
<< SARAH-4.13.0/SARAH.m.
Start["SM"];
\end{MIN}
Afterwards, one can start to play with the unitarity constraints. Here, we want to replace the quartic coupling $\lambda$, which is usually used in the vertices, by the Higgs mass. That's done during the initialisation process of the unitarity constraints. 
\begin{MIN}
InitUnitarity[{Lambda -> mh2/v^2}] 
\end{MIN}
Now, we can take a look at the different scattering processes. The scattering with only the CP even Higgs as external particle is returned by
\begin{MIN}
a0hhhh = Simplify[GetScatteringDiagrams[{hh, hh} -> {hh, hh}]]
\end{MIN}
The result is rather lengthy: \\
\begin{MOUT}
-((3 mh2 (3 mh2 s^2 (tChan + uChan)                                              Log[(s^2-2 s pmass[hh]^2-Sqrt[s^2(s-4 pmass[hh]^2)^2])                        (s^2-2 s pmass[hh]^2+Sqrt[s^2(s-4 pmass[hh]^2)^2])]                + (qChan s + 3 mh2 sChan) Sqrt[s^2 (s - 4 pmass[hh]^2)^2]                    - pmass[hh]^2 (3 mh2 s (tChan + uChan)                                           Log[(s^2-2 s pmass[hh]^2-Sqrt[s^2(s-4 pmass[hh]^2)^2])/                     (s^2-2 s pmass[hh]^2+Sqrt[s^2(s-4 pmass[hh]^2)^2])]             +    qChan Sqrt[s^2 (s - 4 pmass[hh]^2)^2]))) /                           (32 Pi s v^2 (s^2 (s - 4 pmass[hh]^2)^2)^(1/4) (s - pmass[hh]^2)))     
\end{MOUT}
We can simplify it by introducing a short form of the mass and by setting all filters to 1:
\begin{MIN}
Simplify[a0hhhh /. pmass[hh] -> Sqrt[mh2] /. {tChan -> 1, uChan -> 1, qChan -> 1, sChan -> 1}, {s > 0, s > 4 mh2}]
\end{MIN}
The obtained expression is just the one which was already given by Lee, Quigg and Thacker \cite{}: 
\begin{MOUT}
(3 mh2 (-8 mh2^2 - 2 mh2 s + s^2 - 6 mh2 (mh2 - s) Log[mh2/(-3 mh2 + s)]))      /(32 Pi (mh2 - s) Sqrt[s (-4 mh2 + s)] v^2) 
\end{MOUT}
We can now go one step further and calculate the entire scattering matrix. In the case of the SM this is a $10\times 10$ matrix.  
\begin{MIN}
Simplify[BuildScatteringMatrix /. {pmass[hh] -> Sqrt[mh2], pmass[_] -> 0}  /. {tChan -> 1, uChan -> 1, qChan -> 1, sChan -> 1},  {s > 0, s > mh2}]
\end{MIN}
Here, we make here the 
same assumptions as above. Moreover, we set all masses but the one of the CP even Higgs to zero, i.e. we take the limit $m_Z=m_{H^+}=0$. The outcome is:\\
\begin{MOUT}
{{(mh2 ((2 mh2-3s)s + 2 mh2 (mh-s) Log[mh2/(mh2+s)]))/(32Pi s (-mh2+s) v^2),   0, 0, 0, (mh2 (s^(5/2) Sqrt[-4 mh2 + s] + 2 mh2 Sqrt[s^3 (-4 mh2 + s)]      + 2 mh2 s (s-mh2) Log[(-2 mh2 s+s^2 - Sqrt[s^3(-4mh2+s)])/(-2 mh2 s+s^2    + Sqrt[s^3 (-4 mh2+s)])]))/(32Pi (mh2-s) s^(7/4) (-4 mh2+s)^(1/4)v^2), 0, 0, 0, (mh2 s)/(Sqrt[2] (16 mh2 Pi v^2 - 16 Pi s v^2)), 0},          {0, ( mh2 ((mh2 - s)^2 (mh2 + s) + mh2 s^2 (Log[mh2^2/(2 mh2 s - s^2)] +     3 Log[(mh2 s)/(mh2^2 - mh2 s + s^2)])))/(  16 Pi (mh2 - s) s^2 v^2),          0, 0, 0, 0, 0, 0, 0, 0},                                          {... 
\end{MOUT}
where we only have shown the first two rows. In order to see the basis in which the matrix is given, one can check
\begin{MIN}
scatteringPairs
\end{MIN}
which reads in our case 
\begin{MOUT}
{{Ah, Ah}, {Ah, hh}, {Ah, Hp}, {Ah, conj[Hp]}, {hh, hh}, {hh, Hp}, {hh, conj[Hp]}, {Hp, Hp}, {Hp, conj[Hp]}, {conj[Hp], conj[Hp]}}
\end{MOUT}
\subsection{Including the Unitarity constraints in the \SPheno Output}
While it might be helpful to obtain analytical expressions for some specific channels, 
in practice a numerical calculation is often more useful. However, \Mathematica is not
the preferred environment for exhaustive, numerical calculations. Therefore, it was natural 
to extend the existing \SPheno output of \SARAH by the new function. So far, \SARAH is already 
producing {\tt Fortran} source code which can be compiled with \SPheno. This provides the possibility to 
calculate many things for a given model very quickly, e.g. two-loop RGEs, one- and two-loop masses \cite{Goodsell:2014bna,Goodsell:2015ira,Braathen:2017izn},
flavour and precision constraints \cite{Porod:2014xia}, two- and three-body decays at tree-level, loop corrections to 
two-body decays \cite{Goodsell:2017pdq}, and so on. 
\subsubsection{Generating the Fortran code}
The properties of the new spectrum generator based on \SPheno are defined within  \SARAH by using the input file {\tt SPheno.m}. 
{\tt SPheno.m} contains for instance the information about the free input parameters expected from the user, 
the boundary conditions at different scales, choices for involved scales and several other settings. With \SARAH 4.13.0 the following
settings are supported:
\begin{lstlisting}[language=SLHA,title=SPheno.m]
AddTreeLevelUnitarityLimits=True;
\end{lstlisting}
This enables the output of all routines to calculate the tree-level unitarity constraints. By default, this generates the full scattering matrix involving 
all scalar fields in the model which are colourless. In the case that some particles should not be included, they can be explicitly removed via
\begin{lstlisting}[language=SLHA,title=SPheno.m]
RemoveParticlesFromScattering={Se,Sv};
\end{lstlisting}
Here, we have for instance decided not to include charged and neutral sleptons in the case of a supersymmetric model. Once {\tt SPheno.m} for a given model has been edited, 
one can proceed as usual to obtain the source code and compile:
\begin{enumerate}
 \item Run 
\begin{MIN}
MakeSPheno[]
\end{MIN}
 to obtain the source code 
 \item Copy the code to a new \SPheno sub-directory 
 \begin{lstlisting}[]
 > cp -r SARAH-4.13.0/Output/$MODEL/EWSB/SPheno SPheno-4.0.2/$MODEL
 \end{lstlisting}
\item Compile the code 
\begin{lstlisting}[]
 > cd SPheno-4.0.2
 > make Model=$MODEL
\end{lstlisting}
\item Run \SPheno 
\begin{lstlisting}[]
 > ./bin/SPheno-4.0.2 
\end{lstlisting}
\end{enumerate}
For the last step, a Les Houches input file must be provided which includes the numerical values for the input parameters as well as settings for \SPheno. 
\subsubsection{Configuring the unitarity calculations}
If the unitarity constraints are turned on the in the \SPheno output, several new settings in the Les Houches input file are available:
\begin{lstlisting}[language=SLHA,title=LesHouches.in.MODEL]
BLOCK SPhenoInput	 	 # 
 440 1               # Tree-level unitarity constraints (limit s->infinity) 
 441 1               # Full tree-level unitarity constraints 
 442 1000.           # sqrt(s_min)   
 443 2000.           # sqrt(s_max)   
 444 5               # steps   
 445 0               # running   
 445 2               # Cut-Level for T/U poles
\end{lstlisting}
\begin{itemize}
 \item[{\tt 440}]: the tree-level unitarity constraints in the limit of large $\sqrt{s}$ can be turned on/off. Those include only the point interactions 
 \item[{\tt 441}]: the full tree-level calculations including propagator diagrams can be turned on/off.
 \item[{\tt 442}]: the minimal scattering energy $\sqrt{s_{\rm min}}$ is set
 \item[{\tt 443}]: the maximal scattering energy $\sqrt{s_{\rm max}}$ is set
 \item[{\tt 444}]: the number of steps in which \SPheno should vary the scattering energy between $\sqrt{s_{\rm min}}$ and $\sqrt{s_{\rm max}}$ is set. \SPheno will store the maximal eigenvalue. 
 For positive values, a linear distribution is used, for negative values a logarithmic one. 
 \item[{\tt 445}]: RGE running can be included to give an estimate of the higher order corrections
 \item[{\tt 446}]: How shall $t$ and $u$-channel poles be treated:
 \begin{itemize}
  \item[{\tt 0}]: no cut at all
  \item[{\tt 1}]: only the matrix element with a potential pole is dropped
  \item[{\tt 2}]: partial diagonalisation
  \item[{\tt 3}]: entire irreducible sub-matrix is dropped
 \end{itemize}
\end{itemize}

\subsubsection{The \SPheno output}
If the unitarity calculations are switched on, the two new blocks appear in the spectrum file written by \SPheno:
\begin{lstlisting}[language=SLHA,title=SPheno.spc.MODEL]
Block TREELEVELUNITARITY #  
       0    1.00000000E+00  # Tree-level unitarity limits fulfilled or not 
       1    7.32883464E+00  # Maximal scattering eigenvalue 
Block TREELEVELUNITARITYwTRILINEARS #  
       0    1.00000000E+00  # Tree-level unitarity limits fulfilled or not 
       1    1.14400778E+01  # Maximal scattering eigenvalue 
       2    1.92105263E+03  # best scattering energy 
      11    5.00000000E+02  # min scattering energy 
      12    5.00000000E+03  # max scattering energy 
      13    2.00000000E+01  # steps 
\end{lstlisting}
Thus, \SPheno gives two results for the unitarity constrains:
\begin{enumerate}
 \item {\tt TREELEVELUNITARITY}: this block contains the old calculation using only point interactions and the large $s$ limit 
 \item {\tt TREELEVELUNITARITYwTRILINEARS}: this block gives the result for finite $s$ including also propagator diagrams
\end{enumerate}
Both blocks contain the following two elements:
\begin{itemize}
 \item[{\tt 0}]: this is overall result and shows if the point is ruled out ({\tt 0}) or not ({\tt 1}) by the unitarity constraints. The condition for this is that the maximal eigenvalue of the scattering matrix is smaller than $1/2.$
 \item[{\tt 1}]: this entry contains the value of the maximal eigenvalue
\end{itemize}
In addition, the block for the $s$-dependent scattering shows:
\begin{itemize}
 \item[{\tt 2}]: what is the value for $\sqrt{s}$ at which the scattering is maximised
 \item[{\tt 11}--{\tt 13}]: this repeats the input for $\sqrt{s_{\rm min}}$, $\sqrt{s_{\rm max}}$ and the number of steps.
\end{itemize}

\section{Summary}
\label{sec:summary}
We have presented an extension of the \Mathematica package \SARAH to calculate unitarity constraints in BSM models. It is now possible to obtain predictions for the maximal element of the scattering matrix 
in a wide range of models without making use of the large $s$ approximation. We have provided generic expressions for the calculations, along with pedagogical derivations, and clarified some technical issues concerning additional gauge bosons and the choice of gauge. We have briefly shown the importance of these improved constraints in the example of the real singlet extended SM. More detailed 
discussions of the effects of the new constraints in doublet and triplet extensions will be given elsewhere \cite{inprepthdm,inpreptriplet}.

\section*{Acknowledgements}
We thank Marco Sekulla for helpful discussions. FS is supported by ERC Recognition Award ERC-RA-0008 of the Helmholtz Association. MDG acknowledges support from the Agence Nationale de Recherche grant ANR-15-CE31-0002 ``HiggsAutomator'', and the Labex ``Institut Lagrange de Paris'' (ANR-11-IDEX-0004-02,  ANR-10-LABX-63). We would like to thank Sophie Williamson and Manuel Krauss for helpful discussions and collaboration on related topics.

\begin{appendix}
\allowdisplaybreaks

\section{Derivation of the partial wave unitarity constraint}
\label{APP:partialwaves}

In this appendix we will present an elementary derivation of unitarity constraints, retaining finite momentum factors that are less widely known (and absent from e.g. \cite{Kang:2013zba}).

First, we define the $S$-matrix in terms of the interaction matrix $T$ as $S=1+iT$. Then in terms of matrix elements of scattering from (multiparticle) states $a$ with a set of momentum $\{p\}$ to a set of states $b$ with a set of momenta $\{k\}$ we have 
\begin{align}
T_{ba} \equiv {}_{\rm out}\bra \{k,b \} |iT| \{p,a\} \ket_{\rm in} \equiv& i\mathcal{M} ( \{p,a\} \rightarrow \{k,b\}) (2\pi)^4 \delta^4 (\{k\} - \{p\}) \equiv i \mathcal{M}_{ba}(2\pi)^4 \delta^4 (\{k\} - \{p\}),
\end{align}
and so 
\begin{align}
\mathcal{M}^\dagger ( \{k,b\} \rightarrow \{p,a\}) =& \mathcal{M}^* ( \{p,a\} \rightarrow \{k,b\}).
\end{align}
Now $S$ must be a unitary matrix, and so the the constraints from unitarity come from
\begin{align}
S S^\dagger = 1 \longrightarrow T^\dagger T + i(T-T^\dagger) = T T^\dagger + i(T-T^\dagger) = 0.
\label{EQ:simpleSS}\end{align}
Then we insert a complete set of states to evaluate $T^\dagger T$:
\begin{align}
  \bra \{k,b \} |T^\dagger T| \{p,a\} \ket=& \sum_{n} d \Pi_n \bra \{k,b \} |T^\dagger | \{q_n, c_n\} \ket \bra \{q_n, c_n\} | T| \{p,a\} \ket.
\end{align}
Now specialising to the case of $2\rightarrow 2$ scattering, we can rewrite the equation as
\begin{align}
- i (\mathcal{M}_{ba}^{2 \rightarrow 2} - (\mathcal{M}_{ba}^{2 \rightarrow 2})^\dagger ) = \sum_c \frac{1}{2^{\delta_c}}\frac{|\mathbf{p}_c|}{16\pi^2 \sqrt{s}} \int d\Omega \mathcal{M}_{ca}^{2 \rightarrow 2} \overline{\mathcal{M}}_{cb}^{2 \rightarrow 2} +  \underbrace{\sum_{n>2} d \Pi_n d\Omega \mathcal{M}_{ca}^{2 \rightarrow n} \overline{\mathcal{M}}_{cb}^{2 \rightarrow n}}_{\ge 0}. 
\end{align}
Here $\delta_c=0$ if the particles in $c$ are not identical, and $1$ if they are identical, to allow us to keep the same phase space region of integration (otherwise we double count), and $\mathbf{p}_c$ is the three-momentum in the centre of mass frame for the pair $c$.

Now for the partial wave analysis, we define the three-vectors vectors $\mathbf{p}_a, \mathbf{k}_b, \mathbf{p}_c$ to lie along the unit vectors
\begin{align}
  \hat{k}_a =& (1,0,0) \nn\\
  \hat{k}_b =& (z_b, \sin \theta_b, 0) \nn\\
  \hat{k}_c =& (z_c, \sin \theta_c \cos \phi_c, \sin \theta_c \sin \phi_c),
\end{align}
where 
$$ z_b \equiv \cos \theta_b, \qquad z_c \equiv \cos \theta_c.$$
We decompose the matrices into partial waves:
\begin{align}
  \mathcal{M}_{ca} =& 16 \pi \sum (2J + 1) P_J (z_c) \hat{a}_J(s) \nn\\
  \mathcal{M}_{cb} =& 16 \pi \sum (2J + 1) P_J (\hat{k}_b \cdot \hat{k}_c) \hat{a}_J(s) ,
\end{align}
where $P_J$ are the Legendre polynomials, satisfying
\begin{align}
\int_{-1}^1 dz  P_J(z) P_{J'} (z) =& \frac{2}{2J+1} \delta_{JJ'}, \qquad P_0 (z) = 1,
\end{align}
to write
\begin{align}
 -2\pi i(\hat{a}_J - \hat{a}_J^\dagger)^{ba} \le& \sum_c \frac{2^{-\delta_c}|\mathbf{p}_c|}{\sqrt{s}} (2J' + 1) (2J''+1)\int d\phi_c dz_c dz_b P_J (z_b) P_{J'} (z_c) P_{J''} (\hat{k}_b \cdot \hat{k}_c) \hat{a}_{J'} \hat{a}_{J''}.
\end{align}
Next we require the identity
\begin{align}
  P_{J} (\hat{k}_b \cdot \hat{k}_c) =& \frac{4\pi}{2 J + 1} \sum_{m=-J}^J Y_{Jm} (\theta_b, \phi_b) Y_{Jm}^* (\theta_c,\phi_c)
\end{align}
where the spherical harmonics satisfy
\begin{align}
Y_{Jm} \propto e^{im\phi} P_J^m (\cos \theta), \qquad Y_{J0} = \sqrt{\frac{2J +1}{4\pi}} P_J (\cos \theta) .
  \end{align}
In our case we have $\phi_b =0$ so
  \begin{align}
   -2\pi i (\hat{a}_J- \hat{a}_J^\dagger) \le& \sum_c \frac{2^{-\delta_c}|\mathbf{p}_c|}{\sqrt{s}}(2J' + 1) (2J''+1) \int d\phi_c dz_c dz_b P_J (z_b) P_{J'} (z_c)  \hat{a}^{J'}_{ca} \ov{\hat{a}}^{J''}_{cb}  \frac{4\pi}{2 J'' + 1} \nn\\
    & \times \sum_m e^{im\phi} P_{J''}^m (z_b ) P_{J''}^m (z_c )
\end{align}
and thus finally
\begin{align}
- \frac{i}{2} (\hat{a}_J - \hat{a}_J^\dagger)_{ba} \le& \sum_c \frac{2^{-\delta_c}|2\mathbf{p}_c|}{\sqrt{s}} \hat{a}^{J}_{ca} \ov{\hat{a}}^J_{cb}.
  \end{align}
This is true for each partial wave separately.

Now we make the definition:
\begin{align}
a_J^{ba} \equiv & \sqrt{\frac{4 |\mathbf{p}_b|| \mathbf{p}_a|}{2^{\delta_a} 2^{\delta_b} s}} \hat{a}_J^{ba}.
\end{align}
Then we have
\begin{align}
  - \frac{i}{2} (a_J - a_J^\dagger)^{ba} =& a_J^{ca} \ov{a}^{cb}_J = a_J^{cb} \ov{a}^{ca}_J  .
\end{align}
Now, since we could have done this in either order, the matrix $a^J_{ba}$ is normal, and thus both it and $a^\dagger$ can be diagonalised with the same unitary matrix, meaning that we can write for the eigenvalues $(a_J^i)$:
\begin{align}
\mathrm{Im} (a_J^i) \le | a_J^i|^2.
  \end{align}

\section{Scattering amplitudes and partial waves from gauge boson propagators}
\label{APP:GaugeProp}

In this appendix we will clarify the fate of scattering amplitudes among scalars where there is a gauge boson propagator. We neglect all contributions to the final amplitude that are proportional to gauge couplings, but if we work away from the Feynman gauge (it may be desirable to define a theory in that way) then the vector propagators have factors of $1/m_V^2$ where $m_V$ is the vector boson mass -- which is proportional to the gauge couplings. In which case we would necessarily need to included gauge boson amplitudes as well as the Goldstone bosons. 

We write the couplings of a massive gauge boson to real scalars $\phi_i$ as 
\begin{align}
\lagr \supset - \frac{1}{2} g^{V ij} A_\mu^V \phi_i \partial^\mu \phi_j = - \frac{1}{2} \sum_{i > j} g^{V ij} A_\mu^V (\phi_i \partial^\mu \phi_j - \phi_j \partial^\mu \phi_i),
\end{align}
where $V$ now becomes an index.
Then the matrix elements for scalar processes $\{1,2\} \rightarrow \{3,4\}$ considering only the gauge boson propagators are 
\begin{align}
\mathcal{M}^{ba} (\mathrm{Vector\ propagators})=& - g^{V12} g^{V34} \frac{t-u}{s- m_V^2} - g^{V12} g^{V34} \frac{(m_1^2 - m_2^2)(m_3^2 - m_4^2) }{m_V^2} \bigg( \frac{1}{s - m_V^2} - \frac{1}{s - \xi m_V^2} \bigg) \nn\\
&  + \bigg( (2 \leftrightarrow 3),\ (s \leftrightarrow t) \bigg) + \bigg( (2 \leftrightarrow 4),\ (s \leftrightarrow u) \bigg).
\end{align}
To these, we should add the contributions from the Goldstone bosons. In \cite{Goodsell:2017pdq} it was shown that, for scalars coupling to the corresponding golstone boson with the same index $V$
\begin{align}
\lagr \supset - \frac{1}{2} \kappa^{Vij} G_V \phi_i \phi_j 
\end{align}
that the couplings are related by
\begin{align}
\kappa^{Vij} = \frac{m_i^2 - m_j^2 }{m_V} g^{Vij}.
\end{align}
Hence when we add the contribution from the Goldstone bosons, we just obtain
\begin{align}
\mathcal{M}^{ba} (\mathrm{Vector\ propagators+\ Goldstones})=& - g^{V12} g^{V34} \frac{t-u}{s- m_V^2} - \kappa^{V12} \kappa^{V34} \frac{1}{s - m_V^2}  \nn\\
&  + \bigg( (2 \leftrightarrow 3),\ (s \leftrightarrow t) \bigg) + \bigg( (2 \leftrightarrow 4),\ (s \leftrightarrow u) \bigg).
\end{align}
This result is manifestly gauge invariant. 
Setting the gauge couplings $g^{V12}$ to zero, we have a remaining piece which is just equal to the contribution from the Goldstone bosons in Feynman gauge! However, we should note that in other gauges it is necessary to include the gauge boson propagators; for example, if we work in unitary gauge then there are no Goldstone boson propagators!

As an aside, if we want to include the contributions from heavy gauge bosons (i.e. not neglect their couplings), it is simple to perform the angular integrations for these contributions, using:
\begin{align}
M^2 \equiv& m_1^2 + m_2^2 + m_a^2 + m_b^2 = s+t+u\nn\\
\frac{s-u}{t- m_V^2} =& \frac{2s-M^2-m_V^2}{t- m_V^2} -1 \\
\int_{-1}^1 dz \frac{t-u}{s- m_V^2} =& \frac{2}{s-m_V^2} \frac{(m_3^2 - m_4^2)(2s - m_1^2 + m_2^2)}{2s} .
\end{align}
 We find 
\begin{align}
\Delta a_{0} = \frac{1}{16\pi \sqrt{2^{\delta_{12}} 2^{\delta_{34}}}} \bigg\{& \frac{g^{V12} g^{V34}}{s-m_V^2} \sqrt{\frac{4 |\mathbf{p}_1 ||\mathbf{p}_3|}{s}} \frac{(m_3^2 - m_4^2)(2s - m_1^2 + m_2^2)}{2s}\nn\\
 +& g^{V13} g^{V24} \bigg[ (2s - M^2 - m_V^2) f_t(s,m_1^2,m_2^2, m_3^2,m_4^2,m_V^2) - 1\bigg]\nn\\
 +& g^{V14} g^{V23} \bigg[ (2s - M^2 - m_V^2) f_t(s,m_1^2,m_2^2, m_4^2,m_3^2,m_V^2) - 1\bigg]\bigg\}.
\end{align}

\section{Scattering Elements in the real singlet extended SM}
\label{app:singlet}
\begin{align}
a_0(hh\to hh)=&-\frac{3 m_h^2 \left(-8 m_h^4-2 m_h^2 s-6 \left(m_h^4-m_h^2 s\right) \log \left(\frac{m_h^2}{s-3 m_h^2}\right)+s^2\right)}{32 \pi  v^2 \sqrt{s \left(s-4 m_h^2\right)}
   \left(s-m_h^2\right)} \\
a_0(hh\to SS)=& -\frac{\lambda_{HS} \left(2 \lambda_{HS} v^2 \left(s-m_h^2\right) \log \left(\frac{-\sqrt{\left(s-4 m_h^2\right) \left(s-4 m_S^2\right)}-2 m_h^2+s}{\sqrt{\left(s-4 m_h^2\right) \left(s-4
   m_S^2\right)}-2 m_h^2+s}\right)+\left(2 m_h^2+s\right) \sqrt{\left(s-4 m_h^2\right) \left(s-4 m_S^2\right)}\right)}{32 \pi  \left(s-m_h^2\right) \sqrt[4]{s^2 \left(s-4
   m_h^2\right) \left(s-4 m_S^2\right)}} \\
a_0(hS\to hS)=&-\frac{\lambda_{HS}}{16 \pi  s \left(s-m_S^2\right) \sqrt{m_h^4-2 m_h^2
    \left(m_S^2+s\right)+\left(m_S^2-s\right)^2}} \times \nonumber \\
   & \Bigg[-\left(m_h^4-2 m_h^2 \left(m_S^2+s\right)+\left(m_S^2-s\right)^2\right) \left(-\lambda_{HS} v^2+m_S^2-s\right) \nonumber \\
   & +\lambda_{HS} s v^2 \left(s-m_S^2\right)
   \log \left(\frac{m_h^4-2 m_h^2 m_S^2+m_S^4-m_S^2 s}{s \left(2 m_h^2+m_S^2-s\right)}\right)\nonumber \\
   & +3 m_h^2 s \left(s-m_S^2\right) \log \left(\frac{m_h^2
   s}{m_h^4-m_h^2 \left(2 m_S^2+s\right)+\left(m_S^2-s\right)^2}\right)\Bigg]\\
a_0(SS\to SS)=& \frac{\left(4 m_S^2-s\right) \left(\lambda_{HS}^2 v^2-12 \lambda_{S} \left(m_h^2-s\right)\right)+2 \lambda_{HS}^2 v^2 \left(m_h^2-s\right) \log \left(\frac{m_h^2}{m_h^2-4
   m_S^2+s}\right)}{32 \pi  \left(s-m_h^2\right) \sqrt{s \left(s-4 m_S^2\right)}} \\
a_0(hh\to ZZ)=& \frac{2 \left(m_h^6-m_h^4 s\right) \log \left(\frac{-\sqrt{\left(s-4 m_h^2\right) \left(s-4 m_Z^2\right)}-2 m_h^2+s}{\sqrt{\left(s-4 m_h^2\right) \left(s-4 m_Z^2\right)}-2
   m_h^2+s}\right)-m_h^2 \left(2 m_h^2+s\right) \sqrt{\left(s-4 m_h^2\right) \left(s-4 m_Z^2\right)}}{32 \pi  v^2 \left(s-m_h^2\right) \sqrt[4]{s^2 \left(s-4 m_h^2\right) \left(s-4
   m_Z^2\right)}} \\
a_0(hZ\to hZ)=& \frac{m_h^2}{16 \pi  s v^2 \left(s-m_Z^2\right) \sqrt{m_h^4-2 m_h^2 \left(m_Z^2+s\right)+\left(m_Z^2-s\right)^2}} \times \nonumber \\
& \Bigg[m_h^2 s \left(m_Z^2-s\right) \left(\log \left(\frac{\left(m_h^2-m_Z^2\right)^2-m_Z^2 s}{s \left(2 m_h^2+m_Z^2-s\right)}\right)+3 \log \left(\frac{m_h^2
   s}{m_h^4-m_h^2 \left(2 m_Z^2+s\right)+\left(m_Z^2-s\right)^2}\right)\right) \nonumber \\
  &  -\left((m_h-m_Z)^2-s\right) \left((m_h+m_Z)^2-s\right)
   \left(m_h^2-m_Z^2+s\right)\Bigg] \\
a_0(SS\to ZZ)=& \frac{\lambda_{HS} \sqrt{s} \sqrt[4]{\left(s-4 m_S^2\right) \left(s-4 m_Z^2\right)}}{32 \pi  m_h^2-32 \pi  s} \\
a_0(ZZ\to ZZ)=& \frac{m_h^2 \left(\left(2 m_h^2-3 s\right) \left(s-4 m_Z^2\right)+2 \left(m_h^4-m_h^2 s\right) \log \left(\frac{m_h^2}{m_h^2-4 m_Z^2+s}\right)\right)}{32 \pi  v^2
   \left(s-m_h^2\right) \sqrt{s \left(s-4 m_Z^2\right)}} \\
a_0(hh\to WW)=&   \sqrt{2}(a_0(hh\to ZZ) | m_Z \to m_W)\\
a_0(hW\to hW)=& (a_0(hZ\to hZ) | m_Z \to m_W)\\
a_0(SS\to WW)=& \sqrt{2}(a_0(SS\to ZZ) | m_Z \to m_W)\\
a_0(WW\to WW)=& \frac{m_h^2 \left(\left(m_h^2-2 s\right) \left(s-4 m_W^2\right)+\left(m_h^4-m_h^2 s\right) \log \left(\frac{m_h^2}{m_h^2-4 m_W^2+s}\right)\right)}{16 \pi  v^2
   \left(s-m_h^2\right) \sqrt{s \left(s-4 m_W^2\right)}}
\end{align}

\end{appendix}

\bibliographystyle{ArXiv}
\bibliography{lit}

\end{document}